\documentclass[letterpaper]{article} % DO NOT CHANGE THIS
\usepackage{aaai25}  % DO NOT CHANGE THIS
\usepackage{times}  % DO NOT CHANGE THIS
\usepackage{helvet}  % DO NOT CHANGE THIS
\usepackage{courier}  % DO NOT CHANGE THIS
\usepackage[hyphens]{url}  % DO NOT CHANGE THIS
\usepackage{graphicx} % DO NOT CHANGE THIS
\usepackage{natbib}  % DO NOT CHANGE THIS AND DO NOT ADD ANY OPTIONS TO IT
\usepackage{caption} % DO NOT CHANGE THIS AND DO NOT ADD ANY OPTIONS TO IT
\usepackage{algorithm}
\usepackage{algorithmic}
\usepackage{amsmath}
\usepackage{newfloat}
\usepackage{listings}
\DeclareCaptionStyle{ruled}{labelfont=normalfont,labelsep=colon,strut=off} % DO NOT CHANGE THIS
\floatstyle{ruled}
\newfloat{listing}{tb}{lst}{}
\floatname{listing}{Listing}
\usepackage{multirow}
\usepackage{tabularx}
\usepackage{booktabs}
\usepackage{enumitem}
\usepackage{tikz}
\usepackage{longtable}
\usepackage{xcolor}
\usepackage{subcaption}

\lstdefinelanguage{json}{
  float,
  floatplacement=t,
  basicstyle=\ttfamily\small,            
  numbers=none, %left    
  numberstyle=\scriptsize,
  stepnumber=1,
  numbersep=8pt,
  showstringspaces=false,
  breaklines=true,
  frame=single,               
  captionpos=b,  
}

\newcommand*\circled[1]{\tikz[baseline=(myanchor.base)] \node[circle,fill=.,inner sep=1pt] (myanchor) {\color{-.}\bfseries\footnotesize #1};}

\usepackage{framed}

\title{Ethical Classification of Non-Coding Contributions in Open-Source Projects via Large Language Models}
\author{
    Sergio Cobos\textsuperscript{\rm 1}, Javier Luis Cánovas Izquierdo\textsuperscript{\rm 1}\\
}
\affiliations{
    \textsuperscript{\rm 1}Universitat Oberta de Catalunya\\
    Barcelona, Spain\\
    \{scobosga,jcanovasi\}@uoc.edu
} 
 
\begin{document} 

\maketitle

\begin{abstract}
The development of Open-Source Software (OSS) is not only a technical challenge, but also a social one due to the diverse mixture of contributors. To this aim, social-coding platforms, such as GitHub, provide the infrastructure needed to host and develop the code, but also the support for enabling the community's collaboration, which is driven by non-coding contributions, such as issues (i.e., change proposals or bug reports) or comments to existing contributions. As with any other social endeavor, this development process faces ethical challenges, which may put at risk the project's sustainability. To foster a productive and positive environment, OSS projects are increasingly deploying codes of conduct, which define rules to ensure a respectful and inclusive participatory environment, with the Contributor Covenant being the main model to follow. However, monitoring and enforcing these codes of conduct is a challenging task, due to the limitations of current approaches. In this paper, we propose an approach to classify the ethical quality of non-coding contributions in OSS projects by relying on Large Language Models (LLM), a promising technology for text classification tasks. We defined a set of ethical metrics based on the Contributor Covenant and developed a classification approach to assess ethical behavior in OSS non-coding contributions, using prompt engineering to guide the model's output. %We put into practice our approach with a large dataset comprising 225 repositories of the top 5 programming languages in GitHub. 
% Our findings highlight key challenges in classifying and measuring ethical behaviors and provide insights into the adherence to ethical guidelines and codes of conduct within OSS communities.
\end{abstract}
 
\section{Introduction} 
\label{sec:introduction}
Open-Source Software (OSS) is a key part of the infrastructure on which our digital society relies~\cite{eghbal2016roads}.
The development and success of OSS relies on the contributions of a global and diverse network of collaborators, generally scattered around the world and with different profiles. %of a fruitful network of collaborators, generally scattered around the world.
While the progress of any software project mainly relies on coding activities, other non-coding tasks are also present, such as reporting bugs, organizing the development tasks or providing feedback, which enable the participation of different contributor profiles~\cite{DBLP:journals/ese/IzquierdoC22}.
As with any software development process, OSS development is both a technical and social endeavor, where effective collaboration and communication among contributors are key to project success~\cite{DBLP:journals/pacmhci/TrinkenreichGWS20,DBLP:conf/icse/ChengG19} for the project success.

The social dimension in OSS projects is particularly relevant, as project sustainability relies on the success of onboarding new contributions and retaining existing ones~\cite{DBLP:journals/software/SteinmacherTG19}.
To support this, several initiatives have been proposed, such as the definition of onboarding guidelines~\cite{DBLP:conf/sigsoft/CasalnuovoVDF15}, the explicit definition of governance policies~\cite{DBLP:journals/cacm/IzquierdoC23} and the adoption of codes of conduct~\cite{DBLP:conf/wcre/TouraniAS17}, being the last one of the approaches most widely adopted recently. 
A code of conduct defines a set of ethical rules to ensure a respectful and inclusive environment. 
Among existing frameworks, the Contributor Covenant~\cite{ContributorCovenant} stands out as one of the most widely recognized in the computer science field. 

The rules outlined in codes of conduct are mainly addressed to non-coding contributions of OSS projects, which include issues (i.e., change proposals or bug reports for the software system under development) and comments to existing contributions, as these are the main communication means to manage the project development process.
However, ensuring that such non-coding contributions follow the defined ethical rules of the project is not trivial, as contributions have to be reviewed and moderated. 
Even if some approaches have been proposed to facilitate the management of codes of conduct in OSS projects (e.g.,~\citeauthor{icseseis}~\citeyear{icseseis}), there is still a lack of effective methods to classify the ethical quality in OSS projects.

% The classification of ethical quality is a challenging task, where traditional approaches based on keyword matching or regular expressions are not effective
Classifying ethical quality of textual content remains a complex task due to the subtle and context-dependent nature of ethically relevant language~\cite{zhang2018hate}. 
Traditional approaches based on keyword matching or regular expressions tend to fail because they cannot distinguish between nuanced categories~\cite{davidson2017automated}. 
Furthermore, recent work has shown that such approaches are highly vulnerable to adversarial manipulations~\cite{grondahl2018love}. 
The emergence of Large Language Models (LLMs) has opened new opportunities to address this challenge, as they have shown promising results in various Natural Language Processing (NLP) tasks, including text classification~\cite{liu2021pretrain, brown2020language}.

In this paper, we propose an LLM-based approach to classify the ethical quality of non-coding contributions in OSS projects, using prompt engineering to guide the output. 
We focus on contributions that involve issues and comments in OSS projects, and we use the Contributor Covenant as a reference code of conduct, given its relevance in this field.
We therefore define ethical dimensions from the Contributor Covenant, and use them to build a prompt-based classifier.
Our approach is validated on a dataset of more than 2,000 manually labeled contributions, and the results show high classification accuracy.
To illustrate the application of the approach in real-world scenarios, we also present a case study including five diverse OSS projects.

%we later put into practice to measure the ethical quality of non-coding contributions from a set of OSS projects hosted on GitHub, one of the most relevant social-coding platforms.
% The results of our empirical analysis show a high presence of positive ethical dimensions in the contributions of OSS projects, while negative dimensions are less frequent.
% Furthermore, the deployment of codes of conduct reveals having an impact on the ethical quality of the projects' contributions, with a slightly higher presence of positive ethical dimensions.
% We report our learnings and reflections on building the classifier and performing the study, such as the lack of evidence for some ethical dimensions, or the lack of data to train the classifier.
% However, despite their limited occurrence, these negative interactions can pose ethical challenges in projects, highlighting the importance of effective mechanisms for their identification and moderation.
% The results of our empirical analysis show a high presence of positive ethical dimensions in the contributions of OSS projects, but also a significant presence of negative ones, which may lead to ethical issues in the projects.

The rest of the paper is structured as follows.
Section~\ref{sec:background} presents the background.
Section~\ref{sec:approach} details our approach.
% Section~\ref{sec:measurement} describes our study on measuring ethical quality, including the dataset construction and the statistical analysis.
Section~\ref{sec:case} presents a case study with five real-world examples. 
Section~\ref{sec:discussion} provides a discussion of key findings and challenges.
Section~\ref{sec:replicability} presents the replicability package.
Sections~\ref{sec:threats} and \ref{sec:related} discuss threats to validity and related work, respectively.
Finally, Section~\ref{sec:conclusion} concludes the paper. % and outlines future research work.

\section{Background}
\label{sec:background}
In this section, we first provide the necessary background regarding the ethical issues in OSS projects, then we present the current state of practice regarding text classification techniques, and we end the section discussing ethical issues when applying LLM-based classifiers. 

\subsection{Ethical Issues in Open-Source Software}
\label{sec:background:ethical}

\begin{table*}[h]
  \centering
  % \small
  \fontsize{9pt}{10pt}\selectfont
  \begin{tabularx}{0.925\textwidth}{ccl@{\hspace{0.5em}}X}
      \multicolumn{1}{c}{\textsc{Type}} & 
      \multicolumn{1}{c}{\textsc{ID}} & 
      \multicolumn{1}{c}{\textsc{Name}} & 
      \multicolumn{1}{c}{\textsc{Description}} \\
      \toprule
      \multirow{5}{*}{\rotatebox{90}{\textbf{Positive}}} 
        & F1  & Empathy and Kindness             & Demonstrating understanding and compassion towards others. \\
        & F2  & Respect for Differences          & Valuing diverse perspectives and backgrounds. \\
        & F3  & Constructive Feedback            & Providing feedback that is helpful and aimed at improvement. \\
        & F4  & Responsibility and Apology       & Taking responsibility for one's actions and apologizing when necessary. \\
        & F5  & Common Good                      & Acting in ways that benefit the broader community. \\
      \midrule
      \multirow{5}{*}{\rotatebox{90}{\textbf{Negative}}} 
        & F6  & Sexualized Language              & Using language that is inappropriate and sexual in nature. \\
        & F7  & Insulting or Derogatory Comments & Making comments that insult or demean others. \\
        & F8  & Public Harassment                & Engaging in behavior that intimidates or harasses others. \\
        & F9  & Publishing Private Information   & Sharing private information about others without consent. \\
        & \textcolor{gray}{F10} & \textcolor{gray}{Inappropriate Conduct} & \textcolor{gray}{Behaving in a manner that is not suitable in a professional setting.} \\
      \midrule
      \textbf{}  & F11 & No Flag                         & Comments that do not exhibit any ethical behaviors. \\
      \bottomrule
  \end{tabularx}
  \caption{Ethical flags proposed by~\citeauthor{icseseis}~\shortcite{icseseis}.}
  \label{tab:flags}
\end{table*}

As with any other collaborative endeavor, the development of OSS projects faces several ethical challenges, which can be grouped into two main areas: (1) contributor profiles and (2) contributor behavior. 
Regarding contributor profiles, the primary challenges include ensuring diversity and inclusion within the community. 
Biases and exclusionary practices can hinder equitable participation, particularly for underrepresented groups. 
Studies have shown that gender and geographical imbalances persist in many OSS communities, limiting the diversity of perspectives and experiences within projects~\cite{DBLP:journals/sqj/SinghBB22}.  

Contributor behavior also introduces significant challenges, such as power dynamics, where a few individuals may hold disproportionate influence over project decisions.
This situation can lead to governance issues and community conflicts.
Community sustainability relies on effective conflict resolution and governance mechanisms to maintain a healthy and inclusive OSS environment.
To address these concerns, codes of conduct have been widely adopted in OSS communities~\cite{DBLP:conf/wcre/TouraniAS17}.

Over the past decade, several initiatives have aimed to define and enforce ethical guidelines in OSS ecosystems (e.g., Mozilla\footnote{\url{https://www.mozilla.org/en-US/about/governance/policies/participation/}} and Apache\footnote{\url{https://www.apache.org/foundation/policies/conduct}}). 
Among these, the Contributor Covenant~\cite{ContributorCovenant}, currently managed by the Organization for Ethical Source\footnote{\url{https://ethicalsource.dev/}}, has emerged as one of the most widely adopted codes of conduct, including in major projects such as Angular, Kubernetes, and React\footnote{\url{https://www.contributor-covenant.org/adopters/}}. 
The Contributor Covenant outlines key ethical principles for fostering respectful and inclusive collaboration, including behavioral expectations and mechanisms for reporting violations.

Building upon this foundation,~\citeauthor{icseseis}~\shortcite{icseseis} proposed a structured framework to manage codes of conduct and to assess ethical engagement in OSS contributions.
Their approach defines a set of behavioral dimensions, referred to as flags, derived from the Contributor Covenant and grouped into positive, negative, and neutral categories.
Table~\ref{tab:flags} presents these flags.
Flags F1--F5 capture positive behaviors (e.g., empathy, respect for differences or constructive feedback), while F6--F9 reflect negative behaviors (e.g., harassment, insulting or derogatory comments).
For instance, the contributions \emph{``Got it, that makes sense now. Thanks for the clarification!''} and \emph{``I think it's a bit confusing to call both entities variants, so for the purpose of this, I'll call them feature...''} would qualify for F1 and F3, respectively.
On the other hand, contributions such as \emph{``it's just a tool, not some kind of messiah. And don't even get me started on how pretentious this code looks''} or \emph{``I'll be waiting for your next commit, watching over every line of code you write''} would qualify for F7 and F8, respectively.
Although F10 was initially proposed to represent inappropriate conduct, it was excluded in this work due to the high ambiguity involved and low reliability when detecting this category.
We will rely on this set of flags in our approach (i.e., F1--F9), plus an additional flag (i.e., F11) to account for cases where no specific positive or negative flag applies; to classify and measure the ethical quality of contributions in OSS projects. 
An example for F11 would be the contribution \emph{``Mind sharing a codesandbox...?''}.
Note that a contribution may include any combination of positive flags (F1--F5), negative flags (F6--F9), or no flag (F11); and each group is mutually exclusive.
In the following sections, we present both the classification process and the measurement method used to analyze ethical quality in OSS projects.

% By applying them to real-world OSS interactions, our approach extends previous efforts in automated moderation~\cite{Shihab2022, Hsieh2023}, offering an empirical analysis of how ethical engagement manifests in non-coding contributions. 

\subsection{Text Classification with Large Language Models}
\label{sec:background:llms}

Existing text classification approaches often fail as they are generally trained on large, general datasets but lack the specificity needed for the software development field and, in particular, the OSS subfield.
Thus, these limitations hinder their ability to distinguish acceptable interactions from inappropriate behaviors, particularly in sensitive cases like implicit toxicity or gender bias~\cite{sap2019, shah2020predictive}. 
Also, domain-specific challenges, such as technical jargon and abbreviations, further constrain pretrained models' effectiveness~\cite{li2022survey}.

To address these limitations, two approaches can be explored, specifically: (1) application of prompt engineering with LLMs, and (2) use of fine-tuning pre-trained models. 
% The former presents high performance to solve similar problems. 
\citeauthor{brown2020language}~\shortcite{brown2020language} demonstrated that LLMs can perform complex classification tasks with minimal instructions, reducing the need for extensive labeled datasets. 
Furthermore,~\citeauthor{liu2021pretrain}~\shortcite{liu2021pretrain} highlighted the adaptability of prompt-based methods for diverse text classification tasks through well-designed instructions.  
In contrast, fine-tuning approaches require significant computational resources and training time, and pose some risks of overfitting in small or domain-specific datasets like OSS. 
\citeauthor{gao2021making}~\shortcite{gao2021making} emphasized that prompt engineering enhances task performance without sacrificing generalization, and that avoiding fine-tuning is advantageous in low-infrastructure environments. 
\citeauthor{schick2021s}~\shortcite{schick2021s} showed that prompt engineering achieves results comparable to fine-tuning with significantly lower computational costs.
Given the commented benefits of the use of prompt engineering, we adopted this approach for our classifier, specifically employing the GPT-4 Mini model\footnote{We used version GPT-4o-mini-2024-07-18.}, which offers a good balance between computational efficiency and accurate classification performance in NLP tasks.

\subsection{Ethical Issues in LLM-based Classification}
\label{sec:background:ethicalllms}

The use of LLMs to evaluate ethical behavior in OSS communities raises fundamental questions about the delegation of moral judgment to automated systems. 
While LLMs demonstrate significant capabilities in text interpretation and contextual understanding, their integration into domains involving value-laden decisions introduces concerns around accountability, legitimacy, and fairness.

A key concern is the delegation of normative reasoning to systems that lack intentionality or moral agency. 
As highlighted by prior work, ethical assessments require interpretive flexibility and contextual awareness, elements that current LLMs only approximate~\cite{mittelstadt2016ethics, binns2018fairness}. 
In this sense, classifying behaviors as ``constructive'' (e.g., F3) or ``harassing'' (e.g., F8) is not purely a technical task but inherently normative, potentially affecting contributors' reputations and community trust.
Moreover, prior research has shown that algorithmic classification systems often reproduce or amplify existing biases~\cite{sap2019, birhane2023values}. 
LLMs trained on large-scale internet data may inadvertently internalize social hierarchies, leading to disparities in how ethical behavior is recognized or penalized across demographic or linguistic lines. 
This is particularly relevant in OSS communities, where global participation entails diverse norms of communication and expression.

Another critical dimension concerns epistemic opacity. 
The internal mechanisms of deep learning models often preclude transparent justifications for specific classifications, thus complicating processes of accountability and contestation~\cite{burrell2016opacity}. 
In community-driven environments like OSS communities, contributors may reasonably ask for the rationale behind moderation actions, and that is a demand that a black-box classifier may not be able to satisfy.

Despite these challenges, LLM-based approaches can play a supportive role in ethical community governance, as they are used as tools for augmentation rather than replacement of human judgment. 
By surfacing potentially problematic or exemplary behaviors, they can assist maintainers in fostering respectful interactions, especially at scale. 
However, this requires careful prompt design, transparency mechanisms, and avenues for human review to ensure ethical alignment with community values.

Our approach embraces this perspective by relying on a structured set of ethical dimensions grounded in the Contributor Covenant and emphasizing interpretability in our prompt design. 
Furthermore, our validation relies on human-annotated datasets, allowing us to critically assess the classifier limitations and guide its use within OSS contexts as a collaborative, not deterministic, actor in ethical evaluation.

\section{Our Approach}   
\label{sec:approach}
We developed an approach based on the previously presented flags, which encapsulate ethical principles essential for fostering a collaborative and respectful environment, and relying on LLMs to perform the classification according to such flags.
Thus, we devised a multi-label classification model designed to identify specific behavioral flags in the non-coding contributions of OSS projects. 

\subsection{Prompt Guidelines}
Being a prompt-based classifier, special attention must be paid to design the prompt, as it significantly influences the model's performance and reliability in classification tasks~\cite{pesic2023definitions, peng2024incubating, wei2022chain}. 
To ensure robust classification, we considered the following key elements recommended by the literature.

\smallskip
\noindent\textbf{Detailed task description.}
Providing an explicit and detailed task description has been demonstrated to significantly enhance model performance by eliminating potential ambiguities and ensuring consistent task interpretation~\cite{pesic2023definitions}.
In our approach, the description specifically aims at the analysis of non-coding contributions to assign specific ethical flags, including a rationale for facilitating the classification and generating a JSON output.

\smallskip
\noindent\textbf{Comprehensive flag definition.}
Precise and meaningful definitions substantially improve classification performance by providing rich context~\cite{pesic2023definitions, peng2024incubating}.
With this aim, we defined clear and comprehensive criteria for each ethical flag, including positive, negative, and neutral behaviors, plus the corresponding illustrative examples. 

\smallskip 
\noindent\textbf{Logical classification constraints.}
Logical constraints are generally incorporated to prevent contradictory or logically inconsistent outputs.
This is a recommended practice in designing effective prompts for text classification~\cite{pesic2023definitions}. 
For example, in our approach, the neutral flag (i.e., F11) has been defined as mutually exclusive from all other flags, thus ensuring clarity in neutral classification. 
% Additionally, the separation of positive and negative behaviors avoids logical contradictions during classification, a recommended practice in designing effective prompts for text classification~\cite{pesic2023definitions}.

\smallskip
\noindent\textbf{Illustrative examples and structured output format.}
Employing illustrative output examples, particularly in the form of chain-of-thought reasoning, significantly improves model accuracy and interpretability~\cite{wei2022chain, pesic2023definitions}.
Thus we provided clear, representative output examples for each ethical flag following the expected JSON-based format. 
Furthermore, such illustrative examples also facilitate the model's comprehension of classification tasks, thus ensuring consistent interpretation and adaptability. 

\subsection{Prompt Refinement}
To develop an effective classifier for detecting ethical behaviors according to the identified flags, we employed a single-prompt approach using the few-shot learning paradigm.
This configuration allows LLMs to perform complex tasks with a small number of examples, without requiring fine-tuning~\cite{brown2020language}.  
To define the prompt, we followed the guidelines described before and applied an iterative refinement process with the aim of optimizing its performance across all ethical flags.

\begin{figure}[t]
    \centering
    \includegraphics{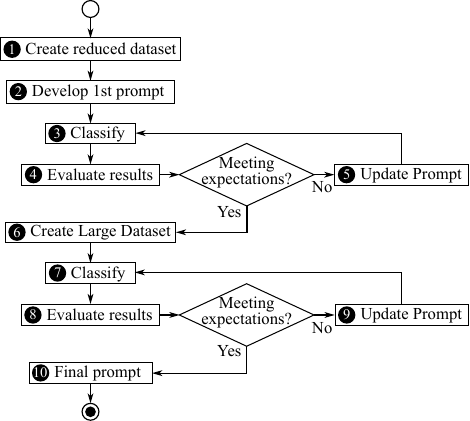} 
    \caption{Prompt Refinement Workflow.}
    \label{fig:workflow-classifier}
\end{figure}

Figure~\ref{fig:workflow-classifier} shows the refinement process followed, which is composed of three main phases, each one using datasets of increasing size.
In the first phase (cf. \circled{1}--\circled{5}), we refined the prompt using a reduced dataset (cf. \circled{1}), containing 50 non-coding contributions from GitHub projects representing all considered ethical flags individually (i.e., 5 contributions per flag).
We selected the GitHub platform as the source of our dataset as it is currently one of the most widely used 
platforms for OSS development, including more than 500 projects, more than 1 million contributors, and over 1 billion contributions\footnote{\url{https://github.blog/news-insights/octoverse/octoverse-2024/}}.
Contributions were randomly collected from the GitHub platform and manually validated to ensure the presence of ethical flags. 
A first prompt was defined (cf. \circled{2}) and iteratively refined (cf. \circled{3}--\circled{5}) until the overall balanced classification precision reached or passed the value of 0.80, thus ensuring rigorousness.
During this process, several refinements were introduced to improve the classifier's performance. 
For example, we explicitly instructed to assign F1 when encountering gratitude expressions (e.g., ``thanks'' or ``thank you''), F3 when any substantial contribution or clarification was made, or F4 when apologies or acknowledgements of fault were present.
Logical constraints were introduced to reinforce internal consistency, as described before.
We also detected that changing the output format (from list to structured JSON) and including additional rationale significantly improved the performance to detect F3 and F11. 
At the end of this phase, we achieved a precision of 0.801.

% During the first phase, we also experimented with multiple prompt formulations (i.e., varying the inclusion of rationales, output format (list vs. structured JSON), and illustrative examples) especially focusing on refining the distinction between F3 and F11, which initially caused frequent confusion. 
% Notably, the format that yielded the best results, with up to 0.15 differences in precision compared to others, was the one using a dictionary-style JSON output with an accompanying rationale per classification decision.

In the second phase (cf. \circled{6}--\circled{9}), we incremented the dataset size to 500 contributions (cf. \circled{6}), where each flag was represented by 50 contributions. 
As before, the contributions were randomly collected from the GitHub platform and manually validated for containing each flag. 
In this phase, the prompt was refined (cf. \circled{7}--\circled{9}) with more detailed definitions and logical constraints. 
Each flag definition was revised to reflect more nuanced interpretations of behavior, including operational criteria and illustrative examples. 
Regarding the positive flags, F1 definition was enriched to include additional expressions of gratitude (e.g., ``thx'', ``thanks for''), thus preventing gratitude contributions from being overlooked or misclassified as neutral; F3 definition was expanded to cover substantial contributions such as error descriptions, solution suggestions, tool recommendations, relevant links, or configuration advice; F4 was adjusted to ensure that any admission of mistakes or explicit apology was correctly labeled; and F5 was reformulated to capture contributions promoting the sustainability or collective well-being of the project. 
Regarding the negative flags, a specific criterion was added to identify F8 when the contribution showed excessive insistence, pressure, or disregard for another person's boundaries, thus reducing misclassifications with F7. 
Furthermore, the prompt was refined to enforce mutual exclusion among flag groups (i.e., positive, negative, and neutral flag groups), thus reinforcing the separation between clearly harmful behaviors and those that, even when critical, maintained an empathetic tone. 
This change improved the model's consistency when handling ambiguous contributions. 
The prompt was refined iteratively until achieving a precision of 0.874, which we considered sufficiently robust to move forward with the large-scale validation and analysis phase, described in the next section. 

The resulting prompt (cf. \circled{10}) is composed of 72 lines of text, including the task description, flag definitions, and illustrative examples\footnote{Due to the lack of space, the complete prompt can be found in the replicability package.}.  

% Finally, the last phase (cf. \circled{10}) involved validating the classifier with a broader set of OSS data, ensuring its applicability beyond the selected samples.  

% The process starts with the first phase and the creation of a reduced dataset (cf. \circled{1}), containing 50 random non-coding contributions from GitHub representing all considered ethical flags. 
% Using this dataset, we designed the initial prompt (cf. \circled{2}) and applied it during the classification stage (cf. \circled{3}). 
% The results of the classification were evaluated (cf. \circled{4}), and iterative modifications were made to the prompt (cf. \circled{5}) to classify correctly each contribution. %with the aim of improving the classification process. 
% We iteratively refined the prompt up to 20 times to address weaknesses, such as improving the recognition of subtle behaviors (e.g., F1 for gratitude and F3 for constructive feedback).
% In the second phase, an expanded dataset of 500 contributions was used (cf. \circled{6}), enabling broader evaluation and further refinements during classification (cf. \circled{7}) and evaluation (cf. \circled{8}). 
% If necessary, the workflow returned to the update stage (cf. \circled{9}) to ensure optimal performance.
% Finally, the classifier was validated with real-world OSS data (cf. \circled{10}). 
% We report the results of this last step in the following section.

\subsection{Dataset for Validation}
% \javi{Extend real vs. synthetic data. Plus tables}

To validate our approach, we built a dataset composed of non-coding contributions from active GitHub public projects.
We consider a project is active if it has more than 1,000 issues with, at least, 300 of them opened; and it shows commit activity in the last 7 days of the data collection.
We randomly selected two projects from GitHub following these criteria, which were \texttt{bmaltais/kohya\_ss} and \texttt{adobe/react-spectrum}.
We collected their non-coding contributions using the GitHub API on January 5$^{th}$, 2025, and built a random sample of 1,000 contributions per project, distributed across issues and comments.

The 2,000 contributions were manually and independently classified by two annotators following predefined guidelines for each ethical flag, reaching a substantial agreement level, as measured by a Cohen's Kappa of 0.79.
This manual classification revealed a very low presence of negative flags (i.e., F6--F9).
% To further investigate whether this scarcity was specific to the selected repositories or indicative of a broader trend, we applied the classifier to an additional set of 20,000 random non-coding contributions from GitHub.
% The results showed a similar distribution of flags, suggesting that the limited presence of negative behaviors is not an isolated case but rather a general pattern. 
To address the scarcity of negative flags in the contributions of our dataset, we created 100 synthetic non-coding contributions for F7, and 50 for each of F6, F8, and F9 (due to overlapping), resulting in a total of 250 additional contributions.
To assist the generation of such a number of synthetic contributions, we relied on a generative AI model\footnote{We used version llama-3-8b-lexi-uncensored.} to kickstart the generation process, and then we manually edited the generated contributions to ensure they accurately represented the intended flag in the OSS context.
Generated contributions were then reviewed by the same annotators, achieving a Cohen's Kappa of~0.94.

All in all, the dataset contained 2,250 non-coding contributions, including all ethical flags. 
To avoid evaluation bias, we included only those with full agreement between the annotators, resulting in a final dataset of 2,024 non-coding contributions. 
This manually validated dataset was used as the ground truth for the evaluation of the classifier.

% To ensure the generalization of the prompt beyond the datasets used in the refinement process, we emphasize that the validation dataset of 2,024 contributions was entirely distinct and independent. 
% The two selected repositories, \texttt{bmaltais/kohya\_ss} and \texttt{adobe/react-spectrum}, were chosen due to their activity and representativeness of diverse interaction dynamics, thus helping to reduce potential specialization or overfitting of the prompt to a single domain. 
% Furthermore, the additional negative contributions were synthetically generated based on randomly selected issue threads from a pool of 100 unrelated repositories, ensuring coverage of a wide variety of topics and community practices. 
% These measures were taken to mitigate potential bias and validate the robustness of the classifier across different OSS contexts.

\subsection{Classifier Performance Evaluation}
This section presents a comprehensive evaluation of the classifier's performance, organized as follows: 
(1) we evaluate the classifier's performance for each ethical flag and (2) grouped by flag type (positive, neutral, and negative); 
(3) we perform a multi-label performance analysis to assess the classifier's capability to predict multiple ethical flags; 
and (4) we examine the classifier's robustness and consistency across multiple executions.

\begin{figure}[t]
    \centering
    \includegraphics[width=\columnwidth]{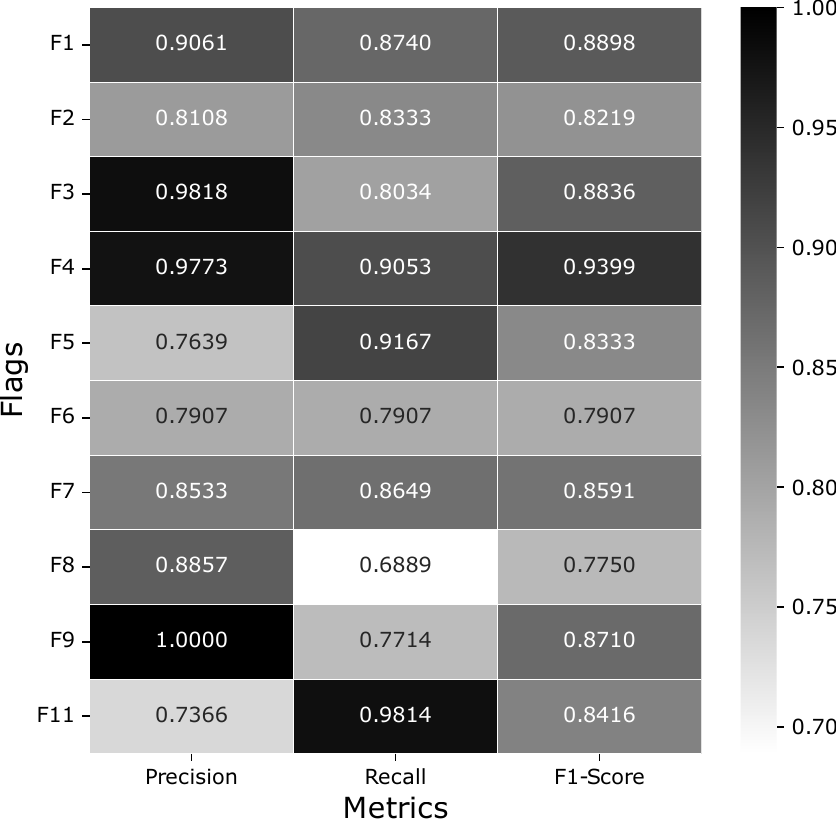}
    \caption{Heatmap of evaluation metrics for each flag.}
    \label{fig:flags_metrics}
\end{figure}

\smallskip
\noindent\textbf{Performance Evaluation per Ethical Flag.} 
We evaluated the classifier's performance to predict individual flags in non-coding contributions. 
Figure~\ref{fig:flags_metrics} shows the results of the analysis. 
As can be seen, the classifier exhibits robust performance in positive categories (i.e., F1--F5), notably excelling in F4, indicating precise and consistent classification with high precision and recall. 
Similarly, F1 shows strong results, although slightly lower recall suggests that some subtle instances may be missed. 
F3 demonstrates %exceptionally 
high precision but reveals occasional difficulty identifying less explicit constructive contributions. 
On the other hand, F5 and F2 display moderate performance, highlighting room for improvement in accurately classifying subtler ethical interactions.

The classifier shows good performance in identifying negative behaviors (i.e., F6--F9), although there is variability between the individual flags. 
F9 achieved the highest performance, driven by excellent precision, though its lower recall suggests difficulties in capturing all relevant instances. 
F7 also shows balanced performance with consistent precision and recall. 
Conversely, F6 displays moderate performance, indicating challenges in accurately capturing subtle or implicit cases. 
Lastly, F8 presents the lowest overall performance due to its relatively low recall, highlighting difficulties in detecting less explicit harassment instances. 
We believe these results are due to the inherent complexity of interpreting patterns of control or intimidation as F8 targets.
% When these nuances were not sufficiently explicit, we detected that the model defaulted to F7. These findings suggest the need for more refined examples and contextual cues to support the distinction between these two negative behaviors.

The neutral flag (i.e., F11) showed moderate results, with high recall but lower precision, indicating a tendency of the classifier to underestimate some ethically relevant contributions and label them as neutral. In particular, we observed false positives in contributions containing minimal but relevant technical content (e.g., links to documentation or testing suggestions), especially when they lacked enough context or intent to meet the criteria of F3.
% This confusion typically arose from the model overestimating the intent to contribute.

% \textcolor{blue}{
% To further understand the limitations of our classifier, we conducted an error analysis focused on the most challenging flags. For F11, we observed that a considerable number of false positives were comments classified as neutral (F11) when they actually contained minimal but meaningful technical content. These often included brief closing remarks such as \textit{“Issue closed”} or confirmations like \textit{“I will test tomorrow”}, which were sometimes mistakenly labeled as constructive feedback (F3). This confusion typically arose from the model overestimating the intent to contribute.
% In the case of F8, most misclassifications were related to F7, as both categories share aggressive or disruptive language. However, F8 specifically targets public harassment, which requires interpreting patterns of control, insistence, or intimidation. When these nuances were not sufficiently explicit, the model defaulted to F7. These findings suggest the need for more refined examples and contextual cues to support the distinction between these two negative behaviors.
% }

\smallskip  
\noindent\textbf{Performance Evaluation per Ethical Flag Type.}  
Previous results allow us to measure the classifier's performance by grouping ethical flags according to their type. 
We therefore group them into positive and neutral flags (F1--F5, F11), and negative flags (F6--F9). 
Table~\ref{tab:all_metrics} summarizes the results obtained from this flag-specific evaluation.

For positive and neutral flags, the classifier demonstrates strong performance, achieving a micro precision of 88.46\% and a micro recall of 86.88\%. 
Micro metrics calculate performance by aggregating all contributions equally, favoring flags with higher occurrences. 
Macro metrics evaluate each flag independently and then average their performance, providing balanced insight across all ethical flags, including those less frequent. 
The consistency of the macro metrics (precision: 88.41\%, recall: 88.57\%) confirms balanced and uniform detection across the positive and neutral ethical flags.
For negative flags, the classifier's performance is slightly lower, with a micro precision of 86.27\% and a micro recall of 81.18\%. 
However, the lower macro recall of 77.90\% emphasizes the model's difficulty to accurately detect subtler or less frequent negative behaviors, underscoring the value of examining both micro and macro metrics for a complete evaluation.

These results demonstrate robust classifier performance, with a global micro precision of 86.82\%, micro recall of 86.25\%, macro precision of 87.06\%, and macro recall of 84.30\%. 
These metrics confirm the classifier's effectiveness while highlighting specific opportunities for improvement, particularly for the detection of subtler negative behaviors.

\begin{table}[t]
    \centering
    \fontsize{9pt}{9pt}\selectfont
    \renewcommand{\arraystretch}{1.1} % Espaciado entre filas
    \setlength{\tabcolsep}{1mm}
    \begin{tabular}{lcc@{\hspace{0.5em}}ccc@{\hspace{0.5em}}ccc}
        \multirow{2}{*}{\textsc{Flags}} & \multicolumn{2}{c}{\textsc{Precision}} & & \multicolumn{2}{c}{\textsc{Recall}} & & \multicolumn{2}{c}{\textsc{F1-Score}} \\
        \cmidrule{2-3} \cmidrule{5-6} \cmidrule{8-9}
        & Micro & Macro & & Micro & Macro & & Micro & Macro \\
        \toprule
        F1--F5, F11 & 0.8846 & 0.8841 & & 0.8688 & 0.8857 & & 0.8766 & 0.8797 \\
        F6--F9      & 0.8627 & 0.8824 & & 0.8118 & 0.7790 & & 0.8365 & 0.8239 \\
        \midrule
        Overall     & 0.8682 & 0.8706 & & 0.8625 & 0.8430 & & 0.8653 & 0.8506 \\
        \bottomrule
    \end{tabular}%
    \caption{Micro and macro performance metrics for positive, neutral, and negative flags.}
    \label{tab:all_metrics}
\end{table}

\smallskip  
\noindent\textbf{Multi-flag Performance Evaluation.} 
% Once assessed the classifier's capability to accurately identify each ethical category individually, 
We also analyzed its performance in a multi-label context, as a single contribution may simultaneously exhibit multiple ethical flags. 
The results show robust performance, with a subset accuracy of 81.12\%, indicating that the classifier correctly predicted all assigned labels simultaneously in the majority of contributions. 
Furthermore, an average precision of 86.90\% suggests a low rate of incorrectly assigned labels, while the average recall of 86.72\% reflects the classifier's strong ability to detect most of the actual labels present in each contribution. 
Finally, an average F1-score of 86.24\% confirms the balanced and consistent performance of the classifier in this multi-label setting.

\smallskip
\noindent\textbf{Classifier Output Consistency and Robustness.} 
To assess the consistency and robustness of the classifier, we conducted three independent runs of the model on the same dataset and compared the resulting outputs. 
This evaluation aims to analyze the stability of the classifier regarding the assigned multi-flags.

We focused on two main metrics: the Exact Match Percentage, which measures the proportion of contributions for which the entire set of flags assigned was completely identical across all three runs, and the Flag Match Percentage, indicating the proportion of contributions were at least one of the assigned flags remained consistent, even if the complete set of flags slightly differed.
The classifier achieved an average Exact Match Percentage of 86.33\% and a Flag Match Percentage of 94.10\%, demonstrating high stability and coherence in its predictions. 
These results highlight the reliability of the model in consistently identifying ethical behaviors across multiple executions on the same dataset.

\smallskip
In summary, the classifier demonstrated a good performance both in identifying individual ethical categories and accurately capturing multiple ethical behaviors. 
It particularly excelled in recognizing positive and neutral behaviors such as responsibility, empathy, and constructive feedback, while facing some challenges in detecting subtler or less frequent negative behaviors, like sexualized language or public harassment.

\section{Case Study}
\label{sec:case}
To showcase the aplication of our approach in real-world escenarios, we present an illustrative case study based on five diverse GitHub projects listed in the first column of Table~\ref{tab:flag_distribution}.
% \texttt{files-community/Files}\footnote{\url{https://github.com/files-community/Files}}, 
% \texttt{immich-app/immich}\footnote{\url{https://github.com/immich-app/immich}}, 
% \texttt{langgenius/dify}\footnote{\url{https://github.com/langgenius/dify}}, 
% \texttt{PowerShell/PowerShell}\footnote{\url{https://github.com/PowerShell/PowerShell}}, and 
% \texttt{microsoft/playwright}\footnote{\url{https://github.com/microsoft/playwright}}.
These projects span a range of domains, governance structures, and community sizes.
We analyzed the distribution of ethical flags in their non-coding contributions during 2024, with the aim to highlight how our approach can reveal meaningful contrasts in community behavior and communication dynamics.
% We believe this case study demonstrates how our approach helps to surface ethical patterns across sociotechnically diverse contexts.
The data collection process was launched on January, 15$^{\text{th}}$ 2025, and included all non-coding contributions published during 2024. 
In total, we collected 7,862 non-coding contributions from the five selected repositories.
Table~\ref{tab:flag_distribution} presents the distribution of flags across these five repositories, serving as the basis for the comparative analysis discussed below.

\begin{table*}[t]
\centering
\fontsize{9pt}{9pt}\selectfont
\renewcommand{\arraystretch}{1.1} % Espaciado entre filas
\small
    \begin{tabular}{lrrrrrrrrrr}
    \toprule
    \textsc{Repository$^{*}$} & 
    \multicolumn{1}{c}{\textsc{F1}} & 
    \multicolumn{1}{c}{\textsc{F2}} & 
    \multicolumn{1}{c}{\textsc{F3}} & 
    \multicolumn{1}{c}{\textsc{F4}} & 
    \multicolumn{1}{c}{\textsc{F5}} & 
    \multicolumn{1}{c}{\textsc{F6}} & 
    \multicolumn{1}{c}{\textsc{F7}} & 
    \multicolumn{1}{c}{\textsc{F8}} & 
    \multicolumn{1}{c}{\textsc{F9}} & 
    \multicolumn{1}{c}{\textsc{F11}} \\
    \midrule
    \texttt{files-community/Files}   & 24.4 &  3.9 & 33.9 & 2.0 &  7.2 & 0.0 & 0.2 & 0.0 & 0.1 & 28.4 \\
    \texttt{PowerShell/PowerShell}   &  7.5 & 12.4 & 43.7 & 1.2 &  8.3 & 0.1 & 2.2 & 0.1 & 0.0 & 24.5 \\
    \texttt{microsoft/playwright}    & 11.9 &  5.2 & 49.0 & 2.0 &  8.1 & 0.0 & 0.7 & 0.2 & 0.1 & 22.8 \\
    \texttt{immich-app/immich}       &  8.5 &  2.6 & 41.6 & 4.6 &  4.4 & 0.0 & 0.4 & 0.0 & 0.0 & 37.8 \\
    \texttt{langgenius/dify}         & 21.8 &  2.9 & 39.2 & 2.0 & 14.6 & 0.0 & 0.6 & 0.1 & 0.1 & 18.5 \\
    \bottomrule
    \multicolumn{11}{l}{{\scriptsize $^{*}$GitHub repository link can be created by adding \texttt{https://github.com/} before the repository name.}} 
\end{tabular}
\caption{Percentage distribution of ethical flags (F1--F11) across selected OSS repositories.}
\label{tab:flag_distribution}
\end{table*}

\smallskip
\noindent \texttt{files-community/Files}. 
This project is focused on enhancing the Windows user experience, and displays a notable presence of empathetic (i.e., F1: 24.4\%) and constructive (i.e.,F3: 33.9\%) contributions.
This is consistent with its development process and strong emphasis on user feedback.
An active Discord server and high responsiveness on GitHub likely contribute to these results. %this collaborative tone.
Additionally, the F5 share (7.2\%) suggests a community concerned with project-wide improvements rather than individual issues.

\smallskip
\noindent \texttt{PowerShell/PowerShell}
Being an official Microsoft-led project, shows high presence of constructive (i.e., F3: 43.7\%) and respectful (i.e., F2: 12.4\%) contributions, but comparatively low empathy (i.e., F1: 7.5\%).
This may reflect its formal governance and the technical profile of its contributors (i.e., primarily system administrators and experienced developers) who often engage in task-focused interactions.

\smallskip
\noindent \texttt{microsoft/playwright}.
This project developes a web testing framework, and shows the highest share of F3 (49.0\%) across all cases, which may reveal its strong emphasis on structured technical feedback.
However, the lower incidence of F1 (11.9\%) and F2 (5.2\%) implies a more transactional or efficiency-driven communication style, which may be aligned with the project's technical purpose and core team leadership.

\smallskip
\noindent \texttt{immich-app/immich}.
This project developes a self-hosted photo management solution, and shows a balanced ethical profile with a significant proportion of F3 (41.6\%) and the highest F11 (37.8\%), suggesting the high presence of contributions without clearly flagged ethical content.
Its F4 level (4.6\%), which indicates acknowledgment of mistakes, is notable for a rapidly evolving project, and may reflect the open, iterative nature of its community practices.

\smallskip
\noindent \texttt{langgenius/dify}.
This repository aims at creating a fast-growing platform for LLM application development, and stands out for its high presence of empathy (i.e., F1: 21.8\%) and community orientation (i.e., F5: 14.6\%).
These results align with its mission to promote inclusive participation and lower entry barriers for AI development, supported by multilingual documentation and accessible support channels such as Discord and Reddit.

\smallskip
While positive behaviors dominate, it is also important to consider negative markers.
In all five projects, the proportion of negative flags (F6--F9) remains low, with no category exceeding 2.2\%.
This suggests generally respectful environments, although such patterns may also reflect the typical reduced expressiveness of technical domains. % or the approach's limitations.

Overall, these contrasting cases illustrate how our approach can uncover distinct ethical communication patterns across OSS projects.
Beyond this exploratory comparison, we believe our approach holds potential for a wide range of applications, including community health monitoring, governance support, early detection of toxic dynamics, and the evaluation of interventions aimed at improving inclusivity and collaboration.
By enabling scalable and interpretable ethical analysis, it offers a novel tool for both researchers and practitioners seeking to better understand and support behavior in collaborative software development.

\section{Discussion} 
\label{sec:discussion}
Beyond the main conclusions reported so far, we would like to highlight some additional insights derived from the development and results of the approach. 

% \smallskip
% \noindent\textbf{Classifier validation and generalization}.
% The classifier's performance directly depends on the validation criteria established during its evaluation. 
% In this work, we were able to perform the validation by employing two evaluators based on agreed-upon criteria. 
% From our experience, sometimes the analysis of ethical behaviors may become controversial, and it is challenging to fully agree on a common ethical classification.
% In this sense, the provision of a larger and more diverse group of experts in ethical behavior within software development contexts may help reduce potential biases stemming from individual interpretations.

\smallskip
\noindent\textbf{Limitations of pretrained models for ethical classification}. 
A key motivation of our approach is the inadequacy of current pretrained text classification models for distinguishing ethical behaviors, a concern that should be considered by practitioners. 
Traditional NLP models are optimized for tasks such as sentiment analysis or toxicity detection, but are insufficient for capturing specific ethical markers, such as expressions of empathy (i.e., F1), respect for differing opinions (i.e., F2), or acknowledgment of responsibility (i.e., F4)~\cite{zhang2018hate, davidson2017automated}. 
Moreover, these models struggle with detecting implicit behaviors, such as sexualized language (i.e., F6) in ambiguous contexts or sarcasm that can alter the intent of a contribution.
Our approach based on prompt engineering using LLMs has shown promising results using our validation dataset, but future work may include comparisons with other models or additional ethically annotated datasets.  

% we employed a prompt engineering approach using LLMs, which has shown promising results using our validation dataset, but future work may include comparisons with other models or additional ethically annotated datasets.  
% Through iterative refinements, we aligned the model's classifications more closely with the predefined ethical categories. 
% While this method improved accuracy, it also introduced challenges related to consistency and reliance on manual adjustments. 
% Future work should explore the development of OSS-specific models trained on ethically annotated datasets or hybrid approaches that integrate prompt engineering with fine-tuning techniques to enhance the reliability and interpretability.

\smallskip
\noindent\textbf{Scarcity of negative behaviors}. 
% Another challenge we faced was the limited presence of negative comments is the wild. 
The scarcity of negative contributions can be attributed to several factors, including repository moderation policies that remove inappropriate content before it is publicly accessible. 
Additionally, contributors in highly visible projects may be more likely to adhere to community guidelines, reducing the prevalence of unethical discourse. 
This is a limitation that impacts the robustness of classification models by restricting their exposure to a diverse range of real behaviors. 
Addressing this issue could involve leveraging alternative data sources, such as archived discussions or external reports of misconduct, to construct a more comprehensive representation of ethical behaviors in OSS. 

\smallskip
\noindent\textbf{Misclassification of constructive feedback as neutral}.
% A particularly notable challenge was the frequent misclassification of neutral comments (i.e., F11, no flag). 
We detected that sometimes contributions initially labeled as neutral were later more accurately identified as constructive feedback (i.e., F3). 
This indicates that the classifier was underestimating the essential contributions being provided.
This classification bias likely stems from the absence of explicit intent markers, causing the model to default to a neutral classification. 
This issue highlighted the inherent difficulty in distinguishing between contributions that are merely neutral and those that provide meaningful insights without explicitly expressing positive sentiment.
We mitigated this issue via prompt refinement, as described in Section~\ref{sec:approach}, incorporating more precise examples that highlighted how non-emotive yet informative contributions help to issue resolution and community discourse. 
Future work could benefit from integrating contextual analysis techniques or supervised learning methods to further reduce false positives and false negatives in ethical classification.

\smallskip
\noindent\textbf{Challenges in evaluating inappropriate conduct}. 
The flag F10 (i.e., conduct which could reasonably be considered inappropriate in a professional setting) was initially considered within the classification framework proposed by~\citeauthor{icseseis}~\shortcite{icseseis}, but it was ultimately excluded in this work due to its broad scope and strong dependence on context.
Unlike other negative behaviors such as sexualized language or insulting contributions, which have well-defined criteria, inappropriate conduct may refer to a wide range of actions whose interpretation usually varies across OSS communities. 
For instance, some projects may enforce strict communication guidelines, while others allow more informal or direct interactions. 
What is considered inappropriate in one setting may be entirely acceptable in another, making it difficult to establish universal classification rules.
Due to this variability, ambiguity, and lack of objective, this ethical category was excluded from the classification process. 
Future research could address this limitation by incorporating adaptive approaches, where community-specific norms define what constitutes inappropriate conduct in each particular context.

\smallskip
\noindent\textbf{Non-coding contributions beyond issues and comments}.
Our empirical study focused on non-coding contributions involving issues and their comments, which we believe account for most of the community discussions on OSS projects. 
In social-coding platforms like GitHub, pull requests are also a way to contribute, which are generally used to submit modifications to the codebase of the project, and therefore could be considered as coding contributions. 
However, pull requests may also trigger some discussions, which may potentially need the application of ethical analysis to ensure that such contributions follow the code of conduct of the project. 
Preliminary analysis revealed that most pull requests follow a template-based format and discussions are very scarce. 
However, we believe further research should better characterize this situation and analyze the ethical behavior in each type of contribution.

\smallskip
\noindent\textbf{Sociotechnical reflections on ethical classification}.
Beyond the technical evaluation, we believe our study raises important sociotechnical considerations. 
Automating the interpretation of ethical behavior in OSS communities via LLMs is not a neutral act, as it involves delegating value judgments to systems that lack moral agency or context. 
As in other fields, in OSS, ethical norms are often defined through lived experience and community negotiation. 
Introducing automated classifiers can shift how norms are enforced and how authority is exercised, potentially reinforcing certain voices while silencing others~\cite{winner1980artifacts, boyd2014critical}.
It is therefore essential to ensure that these systems provide support rather than replace the role of human maintainers and moderators in OSS projects. 
Our approach is aimed at augmenting human judgment, not overriding it, thus functioning as a prioritization tool to assist moderation at scale. 
Preserving human-in-the-loop workflows is crucial to uphold legitimacy and trust in ethical decisions~\cite{green2019algorithmic}.
% Finally, ethical norms evolve. 
Furthermore, a static model, even if well-designed, risks enforcing outdated expectations or failing to detect emergent harms. 
% This highlights the need for adaptive, revisable, and contestable systems, ones that communities can audit, adjust, and align with their shifting values over time~\cite{selbst2019fairness}.
This highlights the need for adaptive systems that communities can audit and align with their shifting values over time~\cite{selbst2019fairness}.
Taken together, these reflections emphasize that the challenge of classifying ethical behavior is not only computational, but also deeply social and political~\cite{binns2018}.

\smallskip
\noindent\textbf{Reflections on annotation diversity}.
% Finally, we must acknowledge a fundamental limitation of our methodology: 
Although all contributions in our evaluation dataset were independently labeled by two annotators, % with substantial agreement, 
ethical interpretation is inherently subjective. 
In many borderline cases, disagreement stemmed not from ambiguity in the flag definitions, but from different assumptions about tone, intent, or social norms in OSS. 
To strengthen future iterations of our approach, %, and its underlying dataset, 
it would be beneficial to include annotators from more diverse backgrounds, particularly those with expertise in social sciences, philosophy, or community moderation. 
Such perspectives could help uncover ethical nuances often overlooked in technically oriented evaluations, and contribute to a more inclusive and reflective approach to ethical automation.

\section{Replicability Package}
\label{sec:replicability}
To facilitate the reproducibility and replicability of our work, we provide a repository\footnote{ZIP file available at \url{https://zenodo.org/records/15494726}} as a companion package for researchers interested in repeating or complementing our study. 
The repository includes the implementation of the classifier, the dataset used for validation, and the results of the performance evaluation.
% The repository includes both the implementation of the classifier and the measurement study. 
% The latter includes the dataset together with the computations and results of the empirical study.

\section{Threats to Validity}
\label{sec:threats} 
Our work is subjected to a number of threats to validity, namely: (1) internal validity, which is related to the inferences we made; and (2) external validity, which discusses the generalization of our findings.
% We present the threats to validity for the classifier approach and the measurement study.

Regarding the internal validity, we relied on data provided by the API of the GitHub platform, which may include low quality non-coding contributions (e.g., incomplete contributions).
To minimize this threat, we focused on two active GitHub projects with large (i.e., higher than 1K) collections of non-coding contributions and manually reviewed the contributions to ensure their quality.
To include contributions with negative flags (i.e., F6--F9), we created 250 additional synthetic contributions using a generative approach, which may include biased elements.
To minimize this threat, two annotators manually validated the generated contributions, adapting them when needed. 
To further investigate whether the scarcity of negative contributions was specific to the selected repositories, we applied our approach to an additional set of 20,000 random non-coding contributions from GitHub.
The results showed a similar distribution of flags, suggesting that the limited presence of negative behaviors is not an isolated case but a general pattern. 

As for the external validity, to ensure the generalization of the prompt beyond the datasets used in the refinement process, we emphasize that the validation dataset of 2,024 contributions was entirely distinct and independent. 
The two selected repositories were chosen due to their activity and representativeness of diverse interaction dynamics, thus helping to reduce potential specialization or overfitting of the prompt to a single domain. 
% Furthermore, the additional negative contributions were synthetically generated based on randomly selected issue threads from a pool of 100 unrelated repositories, ensuring coverage of a wide variety of topics and community practices. 
% These measures were taken to mitigate potential bias and validate the robustness of the classifier across different OSS contexts.
Nonetheless, the results of the performance evaluation should not be generalized to any type of contribution, as it has been specifically validated with non-coding contributions from OSS projects.

% The measurement study is also affected by the same internal threats to validity, as it relies on data from the GitHub API. 
% In this case, to minimize the threat, we focused on the most popular projects developed in the top five programming languages in the platform. 
% Regarding the external validity, our measurement study may be subjected to this threat, as the dataset is based on contributions from a set of GitHub projects for specific programming languages collected on January 5$^{th}$, 2025, and therefore our results should not be directly generalized to other types of OSS projects and programming languages without proper comparison and validation. 

\section{Related Work}
\label{sec:related}
The adoption of codes of conduct in OSS projects has been widely studied, particularly regarding their role in fostering inclusive and respectful communities. 
% Tourani et al.~\cite{DBLP:conf/wcre/TouraniAS17} examined the motivations behind the implementation of codes of conduct in OSS, while Li et al.~\cite{DBLP:journals/pacmhci/LiPFD21} analyzed discussions surrounding the enforcement of codes of conduct on GitHub.
\citeauthor{DBLP:conf/wcre/TouraniAS17}~\shortcite{DBLP:conf/wcre/TouraniAS17} examined the motivations behind the implementation of codes of conduct in OSS, identifying key dimensions such as their purpose, the behaviors they promote, and the enforcement challenges. 
Similarly,~\citeauthor{DBLP:journals/pacmhci/LiPFD21}~\shortcite{DBLP:journals/pacmhci/LiPFD21} analyzed discussions about the enforcement of codes of conduct on GitHub, highlighting tensions between maintainers and contributors in interpreting and applying them. %se guidelines.  

The Contributor Covenant~\cite{ContributorCovenant} has emerged as the most widely adopted code of conduct in OSS, serving as a framework for defining ethical norms in collaborative processes. %development environments. 
\citeauthor{DBLP:journals/sqj/SinghBB22}~\shortcite{DBLP:journals/sqj/SinghBB22} questioned whether codes of conduct genuinely foster ethical engagement or function primarily as symbolic measures with limited practical impact.
While these studies explore the presence and challenges of codes of conduct, our work advances this research by introducing a classification-based approach that enables the systematic measurement of ethical behavior in non-coding contributions. 
% By operationalizing ethical guidelines into a structured set of categories, we provide an empirical analysis of their real-world adherence within OSS communities.  

The classification of ethical behaviors addressed in this work presents unique challenges, as traditional NLP models often struggle to distinguish between neutral, constructive, and unethical interactions.  
For instance,~\citeauthor{renard2024prediction}~\shortcite{renard2024prediction} examined inconsistencies in text classification models, identifying key reliability factors. % that impact their reliability.  
Similarly,~\citeauthor{sap2019}~\shortcite{sap2019} analyzed bias in hate speech detection, raising ethical concerns about automated classification systems and their potential for unfair outcomes.  

Recent advancements in prompt engineering have significantly improved the adaptability of LLMs for domain-specific classification tasks. 
\citeauthor{sahoo2024systematic}~\shortcite{sahoo2024systematic} explored various prompt engineering strategies, demonstrating their effectiveness in guiding model behavior for ethical analysis.  
\citeauthor{shin2023prompt}~\shortcite{shin2023prompt} compared prompting techniques with fine-tuned models, showing that task-specific prompts can %substantially 
enhance classification accuracy.  

Building on these findings, our work employs a structured prompt engineering approach to classify non-coding contributions based on predefined ethical flags.  
Rather than relying on generic sentiment analysis, we adopt a carefully designed framework aligned with the Contributor Covenant, ensuring that classification results accurately capture ethical engagement in OSS interactions.

\section{Conclusion}
\label{sec:conclusion}
In this paper, we have presented an approach to classify the ethical quality of non-coding contributions in OSS projects. 
We rely on a set of ethical categories, or flags, inspired by the Contributor Covenant, %one of the codes of conduct most widely adopted; 
and we leverage LLMs to classify the contributions based on these flags.
Our LLM-based classifier approach shows good performance in the classification of the ethical quality of non-coding contributors.
% We put into practice the classifier by measuring the ethical quality of the non-coding contributions involving the issues and comments of 225 OSS projects hosted on GitHub.
% The results of our empirical study allowed us to better understand the distribution patterns of the ethical flags, and the influence of having explicitly defined the code of conduct in the project.
% We believe our research leads to learnings and reflections which help to pave the road towards an effective and robust analysis of ethical behavior in OSS.

We believe our research introduces a novel approach to classifying and measuring ethical behavior in OSS, and paves the road towards an effective and robust analysis of ethical behavior in other domains using LLMs. 
As our approach is based on prompt engineering, it can easily be adapted to other domains and communities, which may involve prompt-tuning and potentially defining new flags (or adapting the current ones). 
Moreover, since prompt engineering does not require model training or deep infrastructure, this approach can also be adapted and used by individuals without strong technical backgrounds, such as community moderators, ethics scholars, or maintainers, facilitating broader participation in ethical evaluation and governance.

% While previous work has explored the adoption, enforcement, and governance of codes of conduct, our contribution lies in systematically evaluating how these ethical guidelines translate into everyday interactions. 
% By applying prompt engineering techniques, we provide an empirical assessment of ethical engagement in OSS communities, offering new insights into the role of ethical standards in OSS collaboration. 

% However, measuring ethical quality is far from straightforward, as we discuss in Section~\ref{sec:discussion}.
As future work, we are interested in addressing the different lines of work identified in Section~\ref{sec:discussion}.
In particular, we plan to address the scarcity of negative behaviors by exploring additional data sources to build a more representative dataset.
To this aim, we intend to collect data % not only from the GitHub contributions but also 
from other project's resources such as mailing lists or social networks. 
Related to this, we plan to incorporate mechanisms to detect inappropriate conduct (i.e., F10, which was not covered in this work) in OSS projects by adapting the prompt, maybe exploring a more elaborated chain-of-thought. % of performing fine-tuning. 
We are also interested in making more robust the classifier by incorporating knowledge from experts in ethical behavior, who may help in the validation of contributions and improvement of the prompt.
Finally, we plan to explore how the approach performs with non-English languages and apply it to communities beyond OSS projects, which may involve prompt-tuning and defining new flags (or adapting the current ones). 

\section*{Acknowledgement}
This work is part of the project TED2021-130331B-I00 funded by MCIN/AEI/10.13039/501100011033 and European Union NextGenerationEU/PRTR, and the research network RED2022-134647-T.

\bibliography{aies25}

\end{document}